\documentclass[sigconf]{acmart}

\AtBeginDocument{\providecommand\BibTeX{{\normalfont B\kern-0.5em{\scshape i\kern-0.25em b}\kern-0.8em\TeX}}}

\setcopyright{acmlicensed}
\copyrightyear{2024}
\acmYear{2024}
\acmDOI{10.1145/3630744.3663608}

\acmConference[DHOW '24]{Websci Companion ’24, May 21–24, 2024, Stuttgart, Germany}
\acmISBN{979-8-4007-0453-6/24/05}

\usepackage{soul}
\usepackage{xcolor}
\usepackage{subcaption}
\usepackage{amsmath}
\usepackage{amsthm}
\usepackage{booktabs}
\usepackage{comment}
\usepackage{enumitem}
\usepackage{multirow}
\usepackage{tikz}
\usepackage{makecell}

\newcommand\sbullet[1][.5]{\mathbin{\vcenter{\hbox{\scalebox{#1}{$\bullet$}}}}}

\newcommand{\subr}[1]{{\small\texttt{r/#1}}}
\newcommand{\subrtab}[1]{{\footnotesize\texttt{r/#1}}}

\newcommand{\rev}[1]{\textcolor{black}{#1}}

\begin{document}

\title[The Great Ban: Efficacy and Unintended Consequences of a Massive Deplatforming Operation on Reddit]{The Great Ban: Efficacy and Unintended Consequences\\of a Massive Deplatforming Operation on Reddit}
\titlenote{\textcolor{red}{Article published in \textit{Companion Publication of the 16th ACM Web Science Conference}. DOI: \href{https://doi.org/10.1145/3630744.3663608}{10.1145/3630744.3663608}. Please, cite the published version.}}

\author{Lorenzo Cima}
\affiliation{\institution{University of Pisa, IIT-CNR}
  \country{Pisa, Italy}}
  \email{lorenzo.cima@iit.cnr.it}

\author{Amaury Trujillo}
\affiliation{\institution{IIT-CNR}
  \country{Pisa, Italy}}
\email{amaury.trujillo@iit.cnr.it}

\author{Marco Avvenuti}
\affiliation{\institution{University of Pisa}
  \country{Pisa, Italy}}
\email{marco.avvenuti@unipi.it}

\author{Stefano Cresci}
\affiliation{\institution{IIT-CNR}
  \country{Pisa, Italy}}
\email{stefano.cresci@iit.cnr.it}

\renewcommand{\shortauthors}{Cima, et al.}

\begin{abstract}
In the current landscape of online abuses and harms, effective content moderation is necessary to cultivate safe and inclusive online spaces. Yet, the effectiveness of many moderation interventions is still unclear. Here, we assess the effectiveness of The Great Ban, a massive deplatforming operation that affected nearly 2,000 communities on Reddit. By analyzing 16M comments posted by 17K users during 14 months, we provide nuanced results on the effects ---both desired and otherwise--- of the ban. Among our main findings is that 15.6\% of the affected users left Reddit and that those who remained reduced their toxicity by 6.6\% on average. The ban also caused 5\% users to increase their toxicity by more than 70\% of their pre-ban level. Overall, our multifaceted results provide new insights into the efficacy of deplatforming. As such, our findings can inform the development of future moderation interventions and the policing of online platforms.
\end{abstract}



\maketitle

\section{Introduction}
\label{sec:introduction}
Within online platforms, content moderation is needed to maintain safe and inclusive social environments by mitigating the spread of problematic content and harmful behavior. It fosters user trust and safety, thereby upholding ethical standards and contributing to the overall flourishing of healthy online communities~\cite{gillespie2018custodians}. Hence, platform policies are enforced through moderation interventions~\cite{trujillo2023dsa}. There exist many possible interventions to be applied by a platform, ranging from short warning messages~\cite{katsaros2022reconsidering} and the use of informative labels~\cite{zannettou2021won,papakyriakopoulos2022impact} up to the removal of large amounts of content or users~\cite{trujillo2022make,jhaver2021evaluating}. However, in spite of the growing reliance on content moderation, there is still a limited understanding of the general effectiveness of moderation interventions, which impair the efficacy of current regulatory efforts. Indeed, recent research has shown that some interventions yielded mixed~\cite{horta2021platform,trujillo2023one} or no effects at all~\cite{dias2020emphasizing}, and in some cases, some even resulted in unintended and undesired consequences~\cite{bail2018exposure,pennycook2020implied}. For these reasons, it is paramount to conduct thorough assessments of the outcomes of recent moderation interventions as a fundamental preliminary step to the planning and development of future effective solutions.

Out of all the possible interventions, the removal (i.e., banning) of users, communities, and content ---a moderation practice called \textit{deplatforming}--- is by far the most frequently adopted~\cite{ribeiro2024deplatforming}. A famous example is the banishment of Donald Trump from Facebook and X (formerly Twitter) in 2021~\cite{seeliger2023twitter}. Other notable examples are the deplatforming of some toxic influencers on X~\cite{jhaver2021evaluating}, \rev{the removal of accounts involved in coordinated inauthentic behaviors on X~\cite{cima2024coordinated},} and the permanent shut of racist, sexist, and generally hateful communities on Reddit~\cite{chandrasekharan2017you,habib2022proactive}. In 2020, Reddit carried out a massive deplatforming operation that involved the ban of around 2,000 subreddits, on the grounds of their promotion of hate of groups based on identity and vulnerability.\footnote{\url{https://www.reddit.com/r/announcements/comments/hi3oht/update_to_our_content_policy/} (accessed: 03/15/2024)} Among the banned communities were the very popular subreddits \subr{The\_Donald} and \subr{ChapoTrapHouse}, with hundreds of thousands of subscribers. This event is commonly referred to as \textit{The Great Ban} and is currently one of the biggest bans in the history of social media.
Despite its impact on multiple communities and on a large number of users within and beyond Reddit, its effects remain under-explored. Existing studies on The Great Ban have primarily focused on assessing writing style changes~\cite{trujillo2021echo}, which do not answer the fundamental question as to whether the ban was effective at curbing toxic behavior. The few studies that sought to analyze the degree of toxicity after the ban did so only for a few of the affected subreddits~\cite{trujillo2022make}. Moreover, the majority of such studies investigated community-level effects, overlooking user-level reactions that are however clearly relevant with respect to the moderation goal of thwarting particularly problematic users and behaviors~\cite{trujillo2023one}. 

\textbf{Research questions.} Here, we contribute to filling these knowledge gaps by carrying out a large-scale quantitative analysis of the changes in toxic behavior exhibited by those users who participated in the 15 most popular subreddits shut during The Great Ban. We analyze 16M comments shared by almost 17K users over 14 months, answering to the following research questions. 

\noindent$\sbullet[.75]\;$\textbf{RQ1:} \textit{Was The Great Ban effective at reducing toxicity?} The affected subreddits were banned due to hateful and toxic speech. However, previous studies showed that botched interventions can exacerbate ---rather than mitigate--- toxic behaviors. Here we evaluate the effectiveness of The Great Ban at reducing user toxicity. 

\noindent$\sbullet[.75]\;$\textbf{RQ2:} \textit{Did The Great Ban cause any undesired side effects to some users?} In other words, \textit{were there users who became much more toxic after the intervention?} 
When evaluating the outcomes of a moderation action, it is important to consider whether the action made a subset of users resentful, thus leading to marked increases in their toxicity. Such extreme reactions can also occur in the presence of an overall (i.e., platform- or community-level) reduction in total toxicity, which mandates analyses at the \textit{user level}. Here we investigate the presence and quantify the extent of toxic reactions to The Great Ban, as an indicator of potential shortcomings in the intervention. 

\textbf{Main findings.} Based on our answers to the above RQs, our study yields the following novel findings:
\begin{itemize}
    \item Overall, The Great Ban caused 15.6\% of the affected users to abandon Reddit. Those who remained decreased their toxicity by 6.6\%, on average.
    \item Despite this modest overall reduction in toxicity, a non-negligible fraction of users became much more toxic. For example, 5\% users increased their toxicity by more than 70\% of their pre-ban level.
    \item The presence of resentful users who increased their toxicity was widespread across the analyzed subreddits.

\end{itemize}
Overall, our study provides a comprehensive account of the effects of The Great Ban. It surfaces and describes an undesired side effect of the intervention, drawing attention to the delicate balance entailed by the moderation of heterogeneous communities. As such, our results can inform future moderation strategies and the development of effective interventions.
 \section{Related Work}
\label{sec:related-work}
We summarize and critically discuss recent literature on the evaluation of moderation interventions, starting from those works that are most similar to our present study. 

\subsection{Deplatforming}
\label{sub:deplatforming}
Despite the relevance and the extent of The Great Ban, few works delved into a systematic evaluation of its effects. Among them is the study by Milo Trujillo et al.~(\citeyear{trujillo2021echo}) that analyzed activity and linguistic changes in the 15 most popular subreddits affected by the ban. They found that top users suffered the largest decreases in activity and that community response was heterogeneous between subreddits, and even between users of a subreddit~\cite{trujillo2021echo}. Here we complement this work by assessing effects in terms of toxicity, rather than activity and language. Other works evaluated deplatforming effects in a subset of the subreddits affected by The Great Ban, or in other subreddits altogether. \citet{chandrasekharan2017you} and \citet{saleem2018aftermath} examined the repercussions of the bans on \subr{fatpeoplehate} and \subr{coontown}, revealing that a substantial number of users departed from Reddit following the interventions. Among those who stayed, a notable reduction in hate speech was observed. However, they also found a considerable portion of users who migrated to other subreddits, doubling their posting activity~\cite{chandrasekharan2022quarantined}. Instead, \citet{horta2021platform} assessed the impact of deplatforming across multiple platforms, concentrating on the migration of users from banned subreddits to newly established platforms. Their findings indicated a significant decline in user activity on the new platforms. However, they also found a subset of users who increased their toxicity and radicalization. Deplatforming was also studied on platforms other than Reddit. For example,~\citet{mekacher2023systemic} studied ban-induced migrations from Twitter to Gettr, finding that politically polarized users are less toxic on fringe platforms as they are exposed to less out-group interactions. \rev{\citet{cima2024coordinated} analyzed massive bans done by Twitter to counteract coordinated inauthentic behaviors.} Finally,~\citet{jhaver2021evaluating} examined Twitter's ban on multiple toxic influencers, revealing a general reduction in conversations about these figures, accompanied by decreased activity and toxicity among their supporters. Nonetheless, they also found a fraction of users who greatly increased activity and toxicity.

Overall, this body of work shows that moderation interventions frequently cause a combination of desired and undesired effects, and that effects vary between different interventions and types of users. To this end, our work leverages and extends previous knowledge by evaluating the effects of The Great Ban ---a massive, yet essentially unexplored, moderation event--- across the dimension of comment toxicity. Our results provide a picture of the effectiveness of the ban on the 15 most popular affected subreddits.

\begin{table*}[t]
    \small
    \setlength{\tabcolsep}{2pt}
	\centering
	\begin{tabular}{lrcrrcrrcrr}
	    \toprule
		&&& \multicolumn{2}{c}{\texttt{IN-BEFORE}} && \multicolumn{2}{c}{\texttt{OUT-BEFORE}} && \multicolumn{2}{c}{\texttt{OUT-AFTER}} \\
		\cmidrule(lr){4-5} \cmidrule(lr){7-8} \cmidrule(lr){10-11}
        \textbf{subreddit} & \textbf{subscribers} && \textit{core users} & \textit{comments} && \textit{users} & \textit{comments} && \textit{users} & \textit{comments} \\
        \midrule
        \subrtab{chapotraphouse} & 159,185 && 9,295 & 1,368,874 && 9,205 & 3,947,894 && 8,319 & 3,157,462 \\
        \subrtab{the\_donald} & 792,050 && 4,262 & 619,434 && 4,132 & 2,578,026 && 3,145 & 1,434,008\\
        \subrtab{darkhumorandmemes} & 421,506 && 1,632 & 35,561 && 1,617 & 1,246,399 && 1,392 & 689,079\\
        \subrtab{consumeproduct} & 64,937 && 1,730 & 60,073 && 1,719 & 1,209,933 && 1,275 & 594,349\\
        \subrtab{gendercritical} & 64,772 && 1,091 & 94,735 && 1,039 & 511,173 && 706 & 287,877\\
        \subrtab{thenewright} & 41,230 && 729 & 5,792 && 726 & 600,057 && 575 & 308,584\\
        \subrtab{soyboys} & 17,578 && 596 & 5,102 && 594 & 454,659 && 432 & 190,570\\
        \subrtab{shitneoconssay} & 8,701 && 559 & 9,178 && 555 & 338,218 && 384 & 140,619\\
        \subrtab{debatealtright} & 7,381 && 488 & 27,814 && 476 & 274,600 && 328 & 117,281\\
        \subrtab{darkjokecentral} & 185,399 && 316 & 3,214 && 308 & 307,876 && 270 & 179,067\\
        \subrtab{wojak} & 26,816 && 244 & 1,666 && 240 & 210,249 && 170 & 81,142\\
        \subrtab{hatecrimehoaxes} & 20,111 && 189 & 775 && 188 & 185,379 && 143 & 96,457\\
        \subrtab{ccj2} & 11,834 && 150 & 9,785 && 145 & 101,165 && 119 & 63,393\\
        \subrtab{imgoingtohellforthis2} & 47,363 && 93 & 376 && 92 & 74,664 && 72 & 43,018\\
        \subrtab{oandaexclusiveforum} & 2,389 && 60 & 1,313 && 59 & 48,774 && 55 & 35,853\\
        \midrule
        \textit{total (unique)} &&& 16,828 & 2,243,692 && 16,540 & 8,235,086 && 13,963 & 5,592,321\\
		\bottomrule
	\end{tabular}
	\caption{Dataset composition. Rows are ordered by number of active users after the ban. Data in \texttt{IN-BEFORE} is related to user activities within the banned subreddits before the ban took place. Data in \texttt{OUT-BEFORE} and \texttt{OUT-AFTER} are related to user activities outside of the banned subreddits, respectively before and after the ban. \texttt{OUT-BEFORE} contains 288 less users than \texttt{IN-BEFORE}, who had no activity outside of the banned subreddits. \texttt{OUT-AFTER} contains 2,577 less users than \texttt{OUT-BEFORE}, who abandoned the platform after The Great Ban.}
\label{tab:dataset}
\end{table*}
 
\subsection{Soft moderation}
\label{sub:light}
So-called \textit{soft} interventions emerged as an alternative to deplatforming, addressing the concerns about censorship and the loss of free speech that often accompany content and user removals~\cite{zannettou2021won}. Trujillo and Cresci~(\citeyear{trujillo2022make,trujillo2023one}) studied quarantines and restrictions: soft interventions that often precede community bans on Reddit. They studied moderation outcomes on \subr{the\_donald}, revealing a general reduction in activity and toxicity, at the cost of an increased political polarization and decreased factuality of shared news~\cite{trujillo2022make,trujillo2023one}. Reddit's quarantine of \subr{the\_donald} was also studied by~\citet{chandrasekharan2022quarantined} and~\citet{shen2022tale}, who concluded that the intervention did not produce meaningful changes, nor in terms of misogyny and racist comments, neither regarding engagement and internal dynamics. Another type of soft intervention is the attachment of warning labels to disputed posts, whose effects were assessed in terms of perceived credibility and obtained engagement. In detail,~\citet{pennycook2020implied} found that the presence of some posts with warning labels increases the perceived credibility of \textit{all posts} without labels, including false ones not yet debunked. \citet{zannettou2021won} found instead that tweets with warning labels were often replied to, to further debunk the claims. However, this resulted in the flagged tweets circulating more and obtaining more engagement than the undisputed ones. Finally,~\citet{katsaros2022reconsidering} carried out an A/B test on Twitter to evaluate the effectiveness of using warning messages to prompt users who are about to post toxic tweets. Their results show that the intervention was overall effective at reducing the posting of toxic tweets. Nonetheless, a small minority of users edited their tweets to make them more toxic after being exposed to the warning message. \\
The above literature on soft moderation interventions corroborates that on deplatforming, confirming that each intervention can elicit both desired and undesired effects. Overall, this body of work underscores the need for further research to evaluate the impact of little-studied interventions.

 \begin{figure}[t]
        \centering
        \includegraphics[width=\columnwidth]{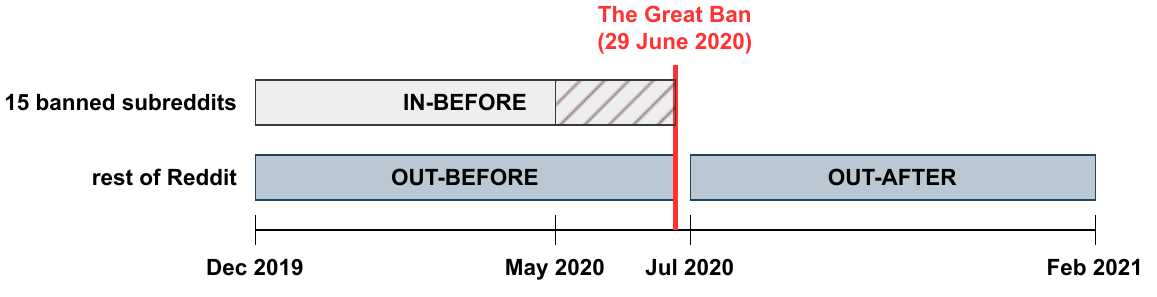}\caption{Timeline depicting the periods of data collection and analysis. Collected data spans two time periods of 7 months each, centered around The Great Ban. It is noteworthy that the \texttt{IN-BEFORE} dataset has no content since May 2020, indicating that the activity in the banned subreddits halted before the official intervention date.}
        \label{fig:period}\end{figure}

\section{Dataset}
\label{sec:dataset}
Our dataset for this study comprises 16M Reddit comments shared by 16,828 distinct users who participated in at least one of the 15 most popular public subreddits in terms of daily active users shut during The Great Ban~\cite{trujillo2021echo}, as reported in Table~\ref{tab:dataset}. 
Even if Reddit administrators originally published an obfuscated list of the most popular banned subreddits,\footnote{\url{https://www.redditstatic.com/banned-subreddits-june-2020.txt} (accessed: 03/15/2024)} previous work deciphered it for those with over 2,000 daily active users~\cite{trujillo2021echo}, resulting in the 15 subreddits used both therein and herein. The composition of our dataset is illustrated in Figure~\ref{fig:period} and the procedure adopted to build it is described in the following.

\textbf{Dataset construction.} We initially collected all comments posted between December 2019 and June 2020 in each of the 15 popular subreddits, resulting in 8M comments shared by 194K distinct users. To collect the comments we used the Pushshift data dumps~\cite{baumgartner2020pushshift} available through Reddit torrents.\footnote{\url{https://academictorrents.com/details/9c263fc85366c1ef8f5bb9da0203f4c8c8db75f4} (accessed: 03/15/2024)} The data covers 30 weeks (i.e., 7 months) prior to The Great Ban, which allows for establishing a suitable baseline for the activity of the affected users before the moderation intervention~\cite{trujillo2023one}. Notably, we could not collect any data between May and June 2020, due to several subreddits having halted their activity, or being banned, before Reddit's public announcement of The Great Ban on June 29, 2020. 

As commonly done in literature, we obtained a representative set of users for the considered subreddits by constraining our analysis to \textit{core users} ---namely, users who participated regularly in at least one subreddit~\cite{trujillo2022make,bouleimen2023dynamics}. We defined core users as those who posted at least one comment each month between December 2019 and March 2020.\footnote{April 2020 was excluded from this analysis due to the limited number of collected comments. It is likely that the activity within the subreddits halted in the initial days of that month.} Additionally, we filtered out bots (i.e., clearly automated accounts) by discarding all accounts that posted at least two different comments at the exact same time.
For this, we used a time delta of at least 1 second between comments, which guarantees we do not inadvertently filter out authentic users~\cite{hurtado2019bot}. Additionally, we manually validated a random sample of 1,000 comments to verify that manifest bots were effectively excluded from the dataset. After these filtering steps, we ended up with 2.2M comments by 16,828 core users. Hereafter, we refer to this portion of our dataset as \texttt{IN-BEFORE} since it involves activity \textit{within} the banned subreddits \textit{before} the ban.

Providing a fair evaluation of the effects of The Great Ban involves matching comparable datasets before and after the intervention. However, no activity exists within the banned subreddits after the intervention, since the subreddits were permanently shut. Therefore, evaluating the effects of the intervention must involve the analysis of user activities \textit{outside} of the banned subreddits. For this reason, we collected all comments made by the 16,828 core users outside of the 15 banned subreddits across a wide time frame spanning 7 months before and after The Great Ban, as shown in Figure~\ref{fig:period}. We obtained around 13.8M comments from 16,540 distinct users. We labeled data related to user activities before the ban as \texttt{OUT-BEFORE} and that after the ban as \texttt{OUT-AFTER}, as reported in Table~\ref{tab:dataset}. Estimates of the effects of the ban are obtained by comparing the \texttt{OUT-BEFORE} and \texttt{OUT-AFTER} datasets. 
Lastly, we enriched our dataset by computing a toxicity score for each collected comment. 

\begin{figure}[t]
    \begin{minipage}[t]{0.49\columnwidth}\centering
        \includegraphics[width=\columnwidth]{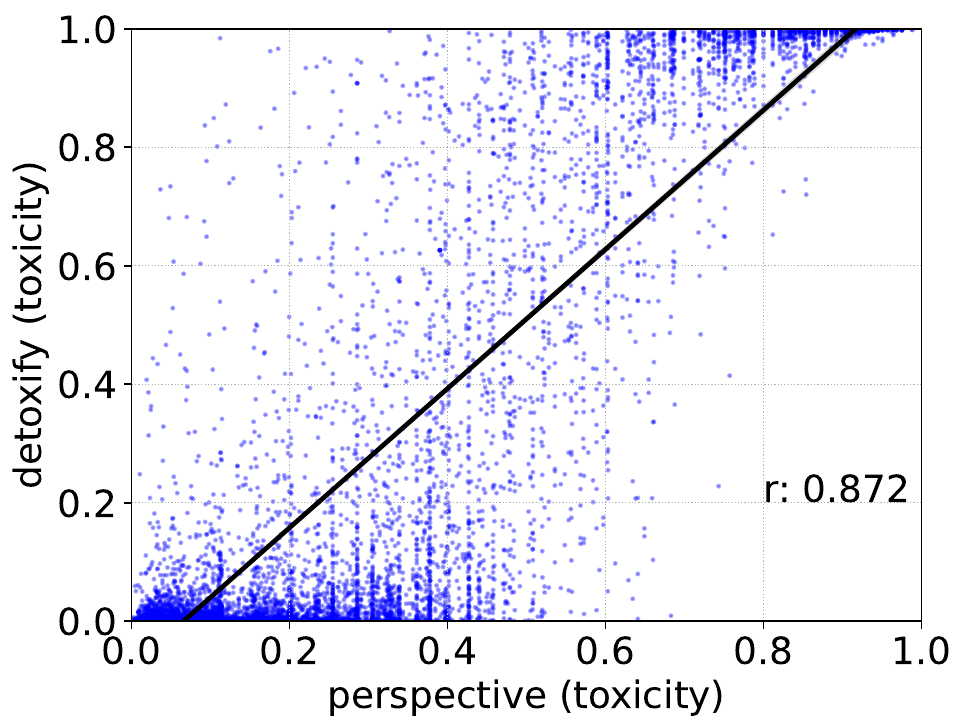}\end{minipage}
    \begin{minipage}[t]{0.49\columnwidth}\centering
        \includegraphics[width=\columnwidth]{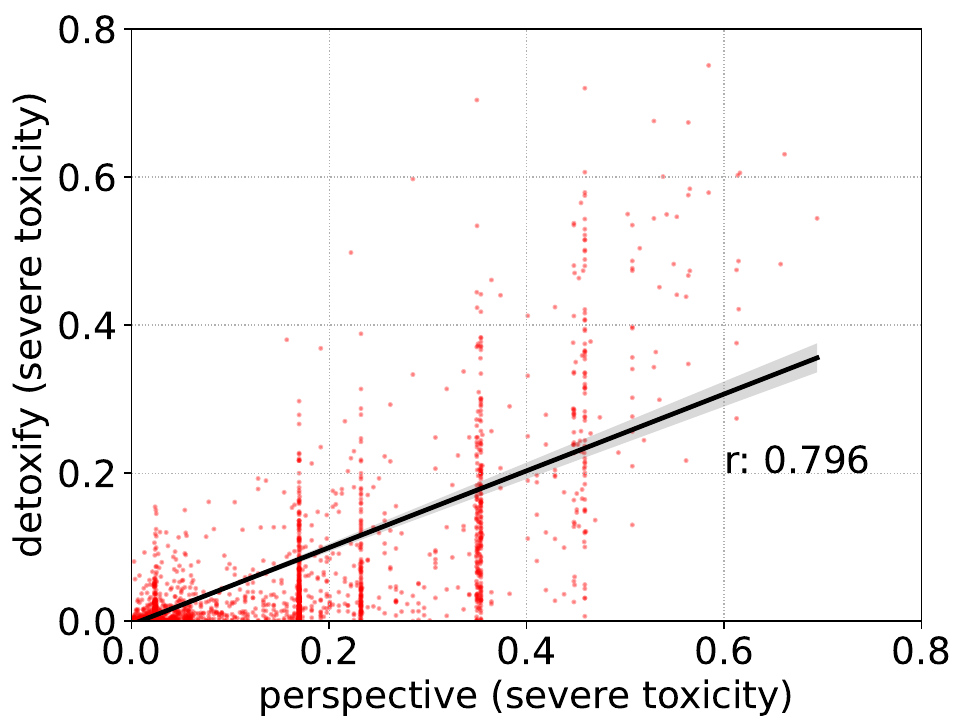}\end{minipage}
    \caption{Comparison between the \textit{toxicity} (left panel) and \textit{severe toxicity} (right panel) indicators provided by Perspective API (\textit{x} axis) and Detoxify (\textit{y} axis). Overall, we report a strong Pearson correlation between the two methods, and particularly so for the \textit{toxicity} indicator ($r = 0.872$).}
    \label{fig:toxicity-comparison}\end{figure}

\textbf{Annotating toxicity: Detoxify \textit{vs} Perspective API.} Google's Perspective API and Detoxify are among the state-of-the-art and most widely used methods for computing toxicity scores for texts. Perspective API\footnote{\url{https://perspectiveapi.com/} (accessed: 03/15/2024)} was developed by the Jigsaw team at Google and currently represents the \textit{de facto} standard for toxicity detection, both in production content moderation environments and in academia~\cite{rieder2021fabrics,nogara2023toxic}. The service is offered as a Web API that, given a piece of text, outputs several scores of offensiveness, including two indicators of \textit{toxicity} and \textit{severe toxicity} defined in the [0, 1] range. Detoxify is an open source deep learning toxicity classifier that also outputs the \textit{toxicity} and \textit{severe toxicity} indicators~\cite{hanu2020detoxify}. Due to its convenience and competitive performance, it has recently seen frequent use~\cite{ejaz2024towards,kopf2024openassistant}. The advantage of Detoxify over Perspective API lies in the possibility of installing and running it locally, without incurring the limitations of a Web API (i.e., rate limits or quotas). Given the large number of comments to annotate in our dataset, we computed toxicity scores with Detoxify. Nonetheless, we first validated our choice by comparing the outputs of Perspective API and Detoxify on a stratified random sample of 10K comments extracted from the \texttt{IN-BEFORE} portion of our dataset. Figure~\ref{fig:toxicity-comparison} presents the results of this comparison, for both the \textit{toxicity} and \textit{severe toxicity} indicators. As shown, we found a strong positive Pearson correlation between the two methods, which supports the use of Detoxify. Then, with respect to the two provided indicators, for our subsequent analyses we relied on the \textit{toxicity} indicator because it is the one on which the two methods agree the most, with $r = 0.872$ \textit{vs} $r = 0.796$ for \textit{severe toxicity}. \section{Analyses and Results}
\label{sec:results}

\begin{table*}[!ht]
    \small
    \setlength{\tabcolsep}{2pt}
	\centering
	\begin{tabular}{lrrrcrrrcrrccccrrcrr}
            \toprule
		& \multicolumn{3}{c} {\textbf{aband. before} (ABA)} && \multicolumn{3}{c} {\textbf{remain. before} (BEF)} && \multicolumn{2}{c} {\textbf{remain. after} (AFT)} && && && && \\
		\cmidrule(lrr){2-4} \cmidrule(lrr){6-8} \cmidrule(lrr){10-11}
        \textbf{subreddit} & \textit{users} & \textit{median} & \textit{MAD} && \textit{users} & \textit{median} & \textit{MAD} && \textit{median} & \makecell[cb]{\textit{MAD}} && \multicolumn{2}{c} {ABA \textbf{\textit{vs}} BEF} && \multicolumn{2}{c} {BEF \textbf{\textit{vs}} AFT} &\\
        \midrule
        \subrtab{chapotraphouse} & 886 & 0.142 & 0.062 && 8,319 & 0.134 & 0.050 && 0.126 & \makecell[cb]{0.052} && $-0.008$ & *** && $-0.008$ & *** \\
        \subrtab{the\_donald} & 987 & 0.148 & 0.072 && 3,145 & 0.134 & 0.051 && 0.127 & \makecell[cb]{0.058} && $-0.014$ & *** && $-0.007$ & *** \\
        \subrtab{darkhumorandmemes} & 225 & 0.169 & 0.050 && 1,392 & 0.152 & 0.044 && 0.142 & \makecell[cb]{0.048} && $-0.017$ & *** && $-0.010$ & *** \\
        \subrtab{consumeproduct} & 444 & 0.179 & 0.053 && 1,275 & 0.158 & 0.045 && 0.146 & \makecell[cb]{0.048} && $-0.021$ & *** && $-0.012$ & *** \\
        \subrtab{gendercritical} & 333 & 0.175 & 0.075 && 706 & 0.145 & 0.061 && 0.122 & \makecell[cb]{0.062} && $-0.030$ & *** && $-0.023$ & *** \\
        \subrtab{thenewright} & 151 & 0.169 & 0.057 && 575 & 0.143 & 0.043 && 0.132 & \makecell[cb]{0.045} && $-0.026$ & *** && $-0.011$ & *** \\
        \subrtab{soyboys} & 162 & 0.194 & 0.050 && 432 & 0.165 & 0.041 && 0.150 & \makecell[cb]{0.044} && $-0.029$ & *** && $-0.015$ & *** \\
        \subrtab{shitneoconssay} & 171 & 0.178 & 0.044 && 384 & 0.157 & 0.044 && 0.146 & \makecell[cb]{0.048} && $-0.021$ & *** && $-0.011$ & *** \\
        \subrtab{debatealtright} & 148 & 0.166 & 0.054 && 328 & 0.149 & 0.047 && 0.139 & \makecell[cb]{0.051} && $-0.017$ & ** && $-0.010$ &  \\
        \subrtab{darkjokecentral} & 38 & 0.190 & 0.040 && 270 & 0.138 & 0.045 && 0.150 & \makecell[cb]{0.039} && $-0.040$ & *** && $-0.012$ & *** \\
        \subrtab{wojak} & 70 & 0.190 & 0.045 && 170 & 0.154 & 0.041 && 0.143 & \makecell[cb]{0.044} && $-0.036$ & *** && $-0.011$ & *** \\
        \subrtab{hatecrimehoaxes} & 45 & 0.177 & 0.042 && 143 & 0.161 & 0.047 && 0.155 & \makecell[cb]{0.058} && $-0.016$ & && $-0.006$ &  \\
        \subrtab{ccj2} & 26 & 0.185 & 0.056 && 119 & 0.123 & 0.036 && 0.114 & \makecell[cb]{0.045} && $-0.062$ & *** && $-0.009$ & * \\
        \subrtab{imgoingtohellforthis2} & 20 & 0.172 & 0.048 && 72 & 0.157 & 0.047 && 0.155 & \makecell[cb]{0.052} && $-0.015$ & && $-0.002$ & \\
        \subrtab{oandaexclusiveforum} & 4 & 0.128 & 0.079 && 55 & 0.189 & 0.064 && 0.182 & \makecell[cb]{0.063} && $+0.061$ & && $-0.007$ & \\
        \midrule
        \textit{overall} & 2,577 & 0.153 & 0.067 && 13,963 & 0.137 & 0.050 && 0.128 & \makecell[cb]{0.053} && $-0.016$ & *** && $-0.009$ & *** \\
		\bottomrule
        \multicolumn{17}{l} {*: $p < 0.1$; **: $p < 0.05$; ***: $p < 0.01$}
	\end{tabular}
	\caption{Subreddit-wise median toxicity scores for users who abandoned Reddit after the ban (ABA) and for those who remained. For the latter, toxicity scores are computed both before (BEF) and after (AFT) the ban. The ABA \textit{vs} BEF and BEF \textit{vs} AFT columns show the effect sizes and the statistical significance levels of the differences in toxicity. The ABA \textit{vs} BEF column shows that users who abandoned the platform were more toxic than those who remained. The BEF \textit{vs} AFT column shows that users who remained active experienced a modest toxicity reduction after the ban.} 
\label{tab:results}
\end{table*}
 
\subsection{RQ1: Effectiveness of The Great Ban}
\label{sec:results-rq1}
In this section, we assess the effectiveness of The Great Ban at reducing toxic behaviors. 

\textbf{Abandoning users.} Table~\ref{tab:dataset} highlights a difference of 2,577 users (15.6\%) between the \texttt{OUT-BEFORE} and \texttt{OUT-AFTER} datasets, corresponding to users who became inactive on Reddit after The Great Ban. Given that such users were consistently active before the ban, but did not post a single comment in the 7 months following it, we conclude that they abandoned Reddit, possibly migrating to other platforms~\cite{horta2021platform}. This result thus highlights a first straightforward effect of The Great Ban. In order to draw more insights into this finding, we compare the pre-ban subreddit-wise toxicity of the abandoning users with that of the users who remained active on the platform. Table~\ref{tab:results} shows that in 14 out of 15 subreddits, users who later abandoned the platform were more toxic than those who remained after the ban. In other words, toxic users were more likely to abandon the platform after the intervention than less toxic ones. For 12 subreddits the difference in toxicity between abandoning and remaining users is statistically significant ($p < 0.05$), according to a non-parametric Mann-Whitney test for unpaired data. Moreover, in 12 subreddits, abandoning users have larger \textit{mean absolute deviation} (\textit{MAD}) scores than remaining users, indicating greater variability in toxicity among the former user group.

\textbf{Remaining users.} In addition to causing some users to abandon the platform, the ban might also have caused toxicity changes in those users who remained. Table~\ref{tab:results} reports subreddit-wise toxicity scores for the matched set of remaining users, before and after the ban. As shown, the remaining users were on average less toxic after the ban. Specifically, when aggregating results for each subreddit, users from all 15 subreddits slightly decreased their toxicity. The decrease is statistically significant for 11 subreddits ($p < 0.1$), according to a non-parametric Wilcoxon test for paired data. Notably, effect sizes for the comparison between remaining users before \textit{vs} after the ban are smaller than those of the comparison between abandoning \textit{vs} remaining users. In fact, after The Great Ban, the overall toxicity decreased by 6.57\% on average ---a modest amount. 
Moreover, for 13 out of 15 subreddits, toxicity \textit{MAD} values are larger after the ban than before. This result suggests that the ban increased the degree of variability in the behavior of the remaining users, which we further investigate in the following.

\begin{figure}[t]
    \centering
    \includegraphics[width=\columnwidth]{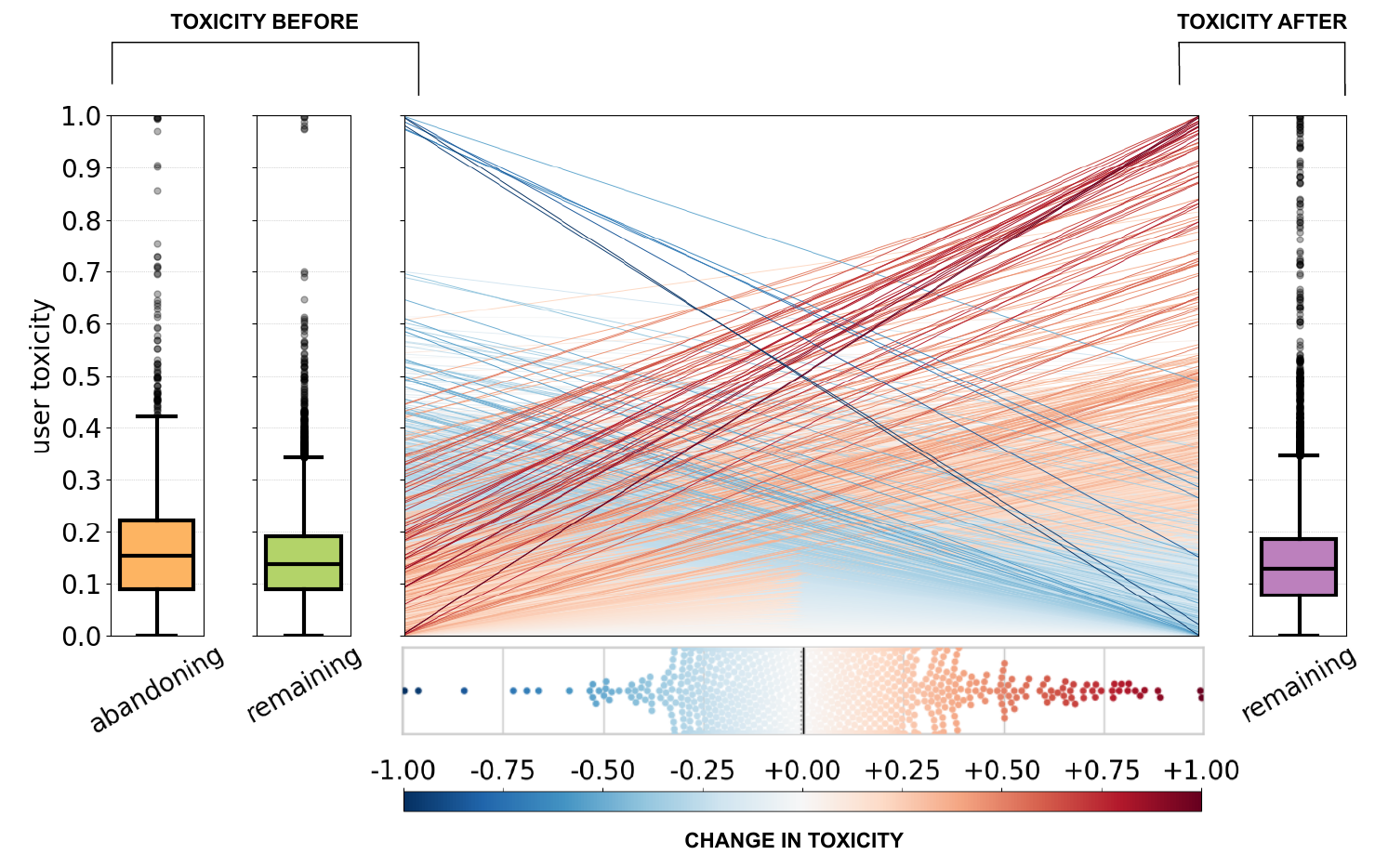}\caption{User-level toxicity change after The Great Ban, for each active user. The slope chart in the central panel reveals a majority of red-colored rising lines, corresponding to a large number of users who drastically increased their toxicity. The beeswarm plot in the bottom panel confirms this effect, presenting more users in the right red-colored tail of the distribution than in the left blue-colored one. Boxplots present marginal distributions for abandoning and remaining user, before and after the intervention.}
    \label{fig:slope-beeswarm-all}\end{figure}

\textbf{User-level effects.} So far, we provided platform- and community-level results, finding that The Great Ban caused 15.6\% of core users to abandon the platform. Furthermore, those who remained exhibited a modest average toxicity reduction of 6.57\%. 
To provide a thorough assessment of the intervention, here we also investigate user-level effects by computing the toxicity changes experienced by each of the 13,963 users who remained active on Reddit after the ban. The central panel of Figure~\ref{fig:slope-beeswarm-all} presents a slope chart of all user-level toxicity changes, independently of the subreddit in which users participated. Each line corresponds to a single user. Line slopes encode the amount of toxicity reduction or increment. Rising lines are colored with different shades of red depending on their slope, and denote users who increased their toxicity after the ban. Contrarily, decreasing lines are blue-colored and denote users who decreased their toxicity. Figure~\ref{fig:slope-beeswarm-all} also includes marginal boxplot toxicity distributions for remaining users before (left-hand side of the slope chart) and after (right-hand side) the ban. The leftmost boxplot presents the toxicity distribution for users who abandoned the platform. Finally, the bottom panel shows the distribution of user-level toxicity changes as a beeswarm plot. This latter plot is useful for highlighting outliers and for studying the two tails of the distribution ---i.e., those related to marked toxicity increases (red dots, right-hand side of the beeswarm plot) and decreases (blue dots, left-hand side).

The three boxplots of Figure~\ref{fig:slope-beeswarm-all} confirm the overall toxicity trends observed in Table~\ref{tab:results}. Abandoning users have the largest median toxicity, followed by remaining users before the ban. Out of the three user groups, the remaining users after the ban have the lowest median toxicity. The slope chart in Figure~\ref{fig:slope-beeswarm-all} depicts a majority of lines with relatively small positive or negative slopes. These correspond to users who exhibited small toxicity changes ---either increases or decreases--- after the ban. At the same time, however, the slope chart also features a remarkable number of steep lines, which are related to users who exhibited large toxicity changes. Specifically, we note an overwhelming majority of red-colored steep lines. This implies that, among users who exhibited large toxicity changes, the vast majority increased their toxicity. This result qualitatively describes an undesired effect of The Great Ban, which caused a non-negligible minority of users to become resentful, thus exhibiting much more toxic behaviors. The points where the lines of the slope chart intersect the \textit{y} axis on the right-hand side of the plot represent the toxicity of the users after the ban. The distribution of the post-ban toxicity scores is also depicted in the right-hand side boxplot. As shown in Figure~\ref{fig:slope-beeswarm-all} and as anticipated from the \textit{MAD} values in Table~\ref{tab:results}, there is more variability in user toxicity after the ban. Figure~\ref{fig:slope-beeswarm-all} clarifies that this increased variability is caused by the increased toxicity of the resentful users.

\begin{figure}[t]
    \centering
    \includegraphics[width=\columnwidth]{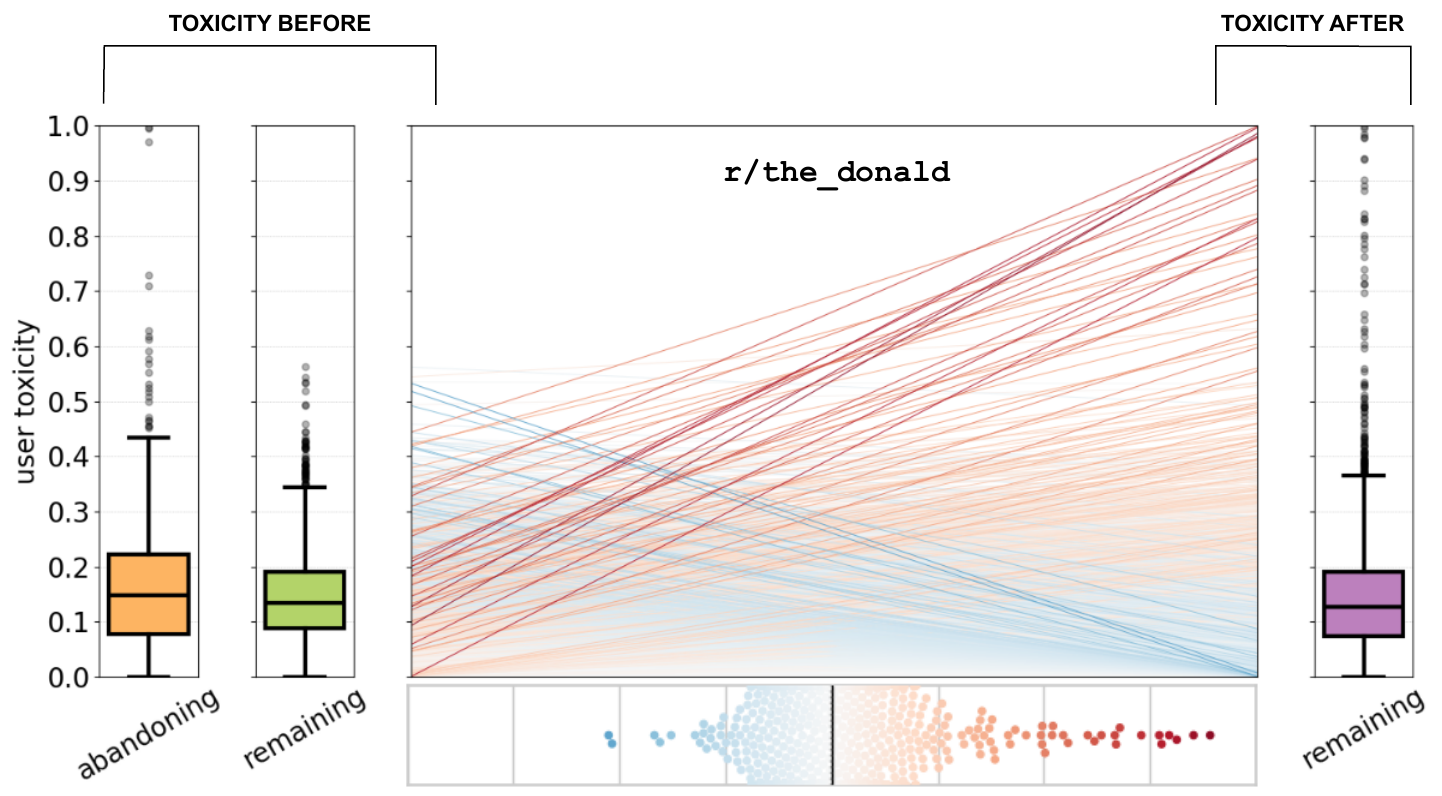}
    \includegraphics[width=\columnwidth]{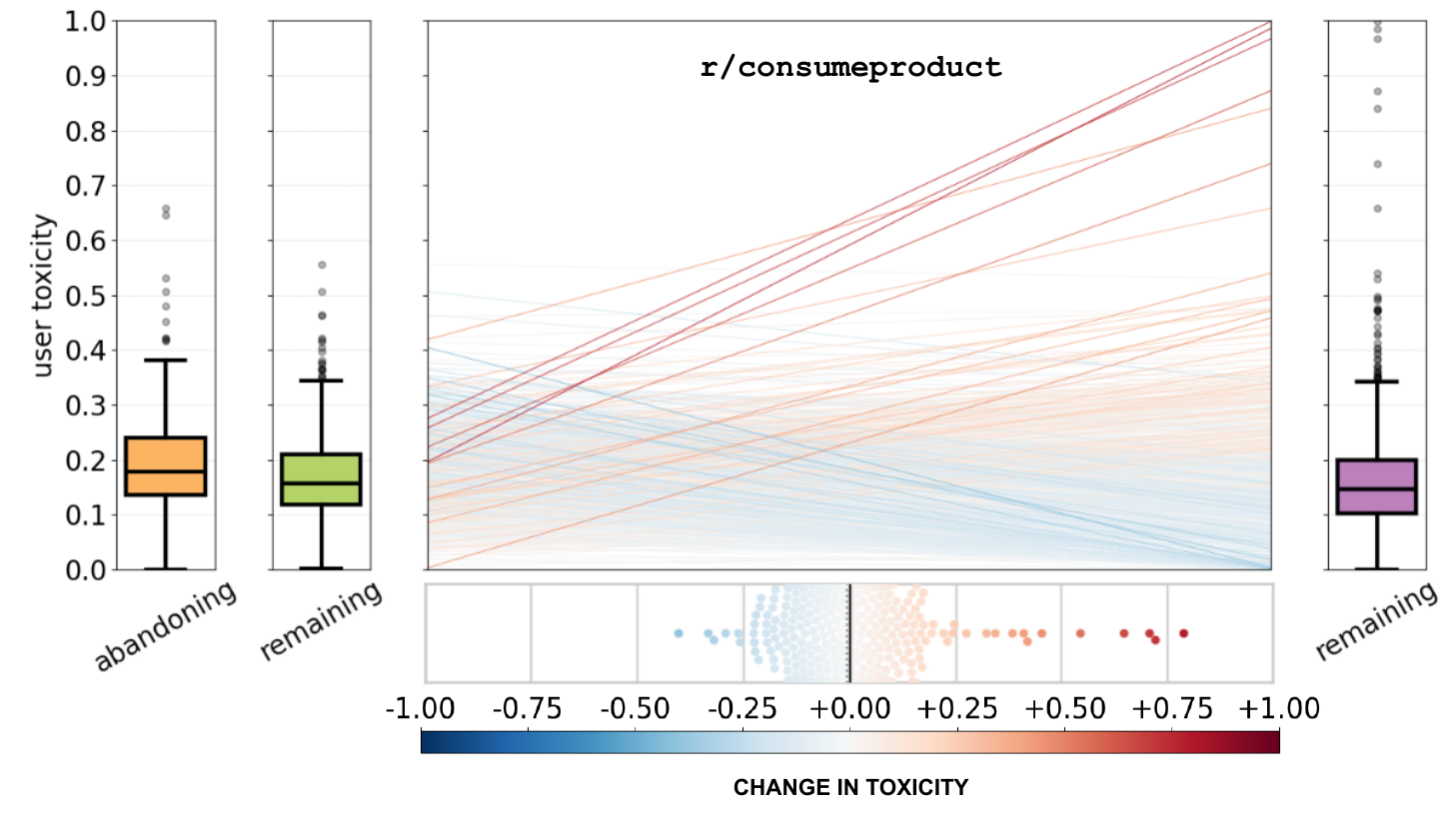}
    \caption{User-level toxicity changes for participants in \subr{the\_donald} (top) and \subr{consumeproduct} (bottom).}
    \label{fig:slope-beeswarm-donald-consume}\end{figure}

Figure~\ref{fig:slope-beeswarm-all} provides aggregated results for all subreddits. However, the same set of visualizations can also be used to assess the consequences of The Great Ban among the users of a single subreddit. In turn, this is valuable for identifying common patterns and possible differences between the subreddits. We thus repeated the analysis for users of each individual subreddit. 
The comparison between the different subreddits confirmed previous results, according to which the majority of users who underwent substantial toxicity changes after the ban, increased their toxicity. Nonetheless, this behavior was more pronounced in certain subreddits while lacking in others. Among the subreddits where the effect was particularly pronounced are \subr{the\_donald} and \subr{consumeproduct}, as visible from the large majority of steep red lines in the slope charts of Figure~\ref{fig:slope-beeswarm-donald-consume}, and from the long right tails of the corresponding beeswarm plots. Conversely, \subr{soyboys} and \subr{hatecrimehoaxes} are subreddits whose participants did not experience marked toxicity changes and for which no extremely toxic behavior was measured post-ban.

\subsection{RQ2: Extreme user reactions to The Great Ban}
\label{sec:results-rq2}
In answering RQ1 we found that The Great Ban caused a modest overall reduction in toxicity. At the same time, however, we also qualitatively discovered a fraction of users who became resentful and greatly increased their toxicity after the ban. 
We are now interested in quantitatively assessing the extent of this issue across the different subreddits.

\begin{figure*}[t]
    \centering
    \includegraphics[width=\textwidth]{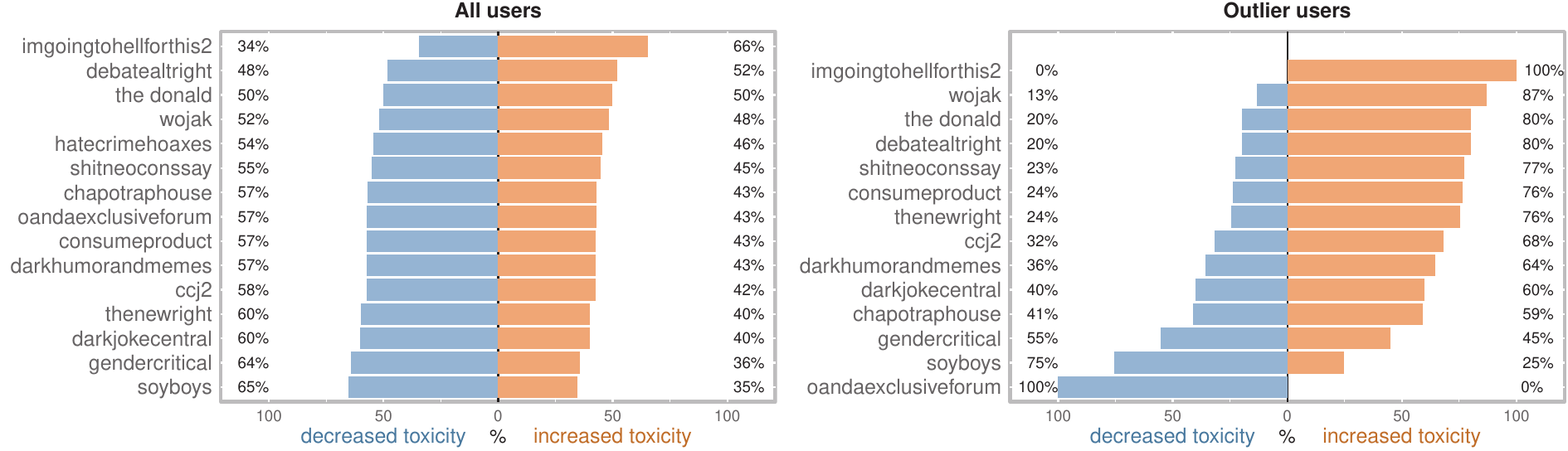}\caption{Subreddit-wise changes in toxicity obtained by summing all individual decreasing and increasing contributions, for all users (left panel) and outlier users (right panel). Outlier users are those whose change in toxicity exceeds the threshold $k = 0.25$. Based on this criterion, we found no outliers for \subr{hatecrimehoaxes}, which is thus omitted from the right panel figure.} 
    \label{fig:all_outliers}\end{figure*}

Let $t(i)_{\text{BEF}}, t(i)_{\text{AFT}}$ be the toxicity of the $i$-th user before and after the ban. Then, $\Delta t(i) = t(i)_{\text{AFT}} - t(i)_{\text{BEF}}$ is the change in toxicity for the same user. For clarity, the beeswarm plot in Figure~\ref{fig:slope-beeswarm-all} shows the distribution of $\Delta t(i)$ for all users. In detail, we measured that 5\% of all users exhibited a $\Delta t(i) > 0.1$. This result is particularly relevant in light of the median toxicity pre-ban, which was 0.137 as reported in Table~\ref{tab:results}. In other words, this result implies that 5\% of users increased their toxicity by more than 70\% of their pre-ban level. 
The overall change in toxicity in a given subreddit with $N$ participants can be computed as $\Delta t = \sum_{i = 1}^N \Delta t(i)$. Then, in order to separately weigh the contribution of the two tails of the beeswarm plots, we compute summations that only consider positive or negative $\Delta t(i)$. For example, we quantify the contribution of the right tail (i.e., the one related to toxicity increases) of the distribution of toxicity changes in a subreddit as:
\[
\Delta t^+ = \sum\nolimits_{i = 1}^N \Delta t(i)\quad \text{with } \Delta t(i) > 0
\]
Similarly, we quantify the contribution of the left tail as $\Delta t^- = \sum_{i = 1}^N |\Delta t(i)|$, with $\Delta t(i) < 0$. The left panel of Figure~\ref{fig:all_outliers} shows the balance between the contributions to the overall toxicity change in each subreddit brought by users who increased ($\Delta t^+$, red-colored) \textit{vs} those who decreased ($\Delta t^-$, blue-colored) their toxicity. As shown, the contributions of the two tails are relatively balanced for the majority of subreddits. In 12 out of 15 subreddits, the decreases in toxicity slightly outweigh the increases, which results in the modest overall decrease in toxicity that we already noted in RQ1. Toxicity increases outweigh decreases in \subr{debatealtright} and \subr{imgoingtohellforthis2}, while \subr{the\_donald} has perfectly balanced contributions.

\begin{figure}[t]
    \centering
    \includegraphics[width=\columnwidth]{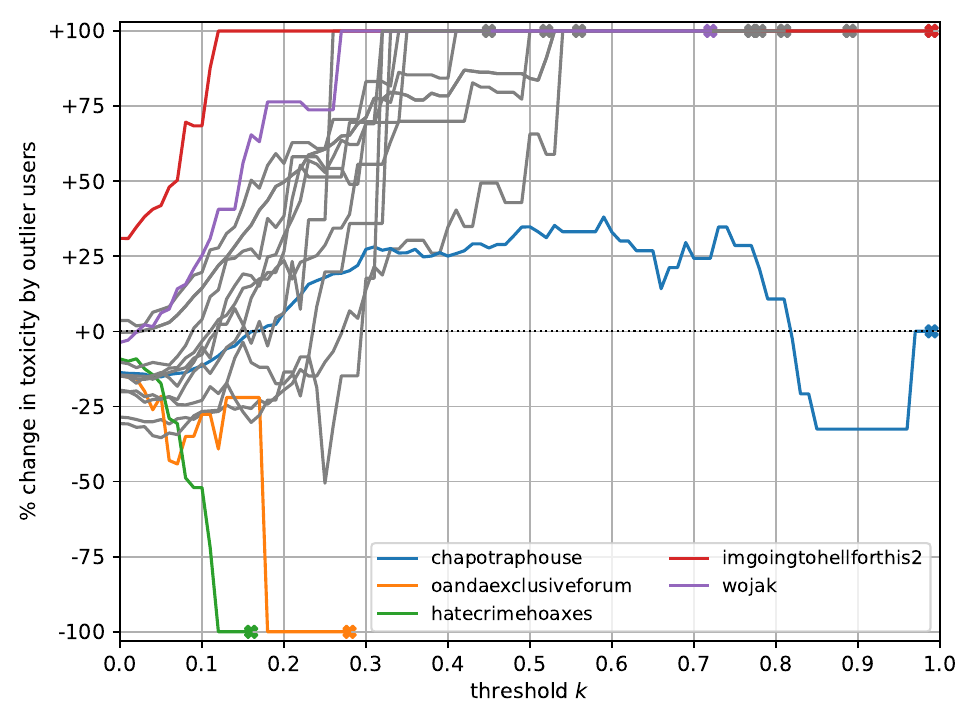}\caption{Contribution of the outlier users to the increase/decrease of toxicity in each subreddit. Outlier users are defined as those whose individual change in toxicity exceeds the threshold $k$. As shown, in 12 out of 15 subreddits the outlier users caused large toxicity increases.} 
    \label{fig:variation}\end{figure}

In order to specifically investigate the behavior of outlier users ---those who experienced marked changes in toxicity--- we recompute the previous summations by only considering users whose change in toxicity exceeds a given threshold $k$: $|\Delta t(i)| > k$. This allows us to focus on those users who experienced extreme behavioral changes, be them increases or decreases in toxicity. For example, the right panel of Figure~\ref{fig:all_outliers} repeats the previous analysis by applying the threshold $k = 0.25$. As shown, we obtain strikingly different results. When only considering outlier users, in 13 out of 15 subreddits the contributions of those who increased or decreased their toxicity to the overall subreddit toxicity are very imbalanced. The only exceptions are \subr{gendercritical} who presents relatively balanced contributions (55\% \textit{vs} 45\%) and \subr{hatecrimehoaxes} for which no user experienced a change in toxicity exceeding the threshold $k = 0.25$. Moreover, in 11 subreddits toxicity increases greatly outweigh decreases. To generalize these findings, we repeated the analysis by varying the value of $k$ from 0 to 1, with a step of 0.01. Results are shown in Figure~\ref{fig:variation} where each line corresponds to a subreddit and reports the \% difference $\Delta t^+ - \Delta t^-$ between the contributions of the two tails, for each value of $k$. In the figure, only a few notable subreddits are highlighted, while the rest are grey-colored so as to reveal the overall trend. As shown, even for small values of the threshold $k$, the vast majority of lines rise steeply. In particular, in 12 out of 15 subreddits the outlier users exhibit large increases in toxicity. The subreddits \subr{imgoingtohellforthis2} and \subr{wojak} are those featuring the largest toxicity contributions by outlier users. On the contrary, outliers in \subr{hatecrimehoaxes} and \subr{oandaexclusiveforum} show marked toxicity reductions, while \subr{chapotraphouse} remains overall stationary. However, the spike observed in \subr{imgoingtohellforthis2}, as well as the decline of \subr{oandaexclusiveforum}, may be attributed to the relatively small number of users (less than 100). In summary, the results presented in this section highlight that the presence of resentful users who became much more toxic after The Great Ban was not localized within a single or a few subreddits. On the contrary, despite the fact that only a minority of users experienced marked toxicity increases, such adverse reactions were pervasive across most of the analyzed subreddits, which is indicative of a systemic phenomenon. 
 \section{Discussion}
\label{sec:discussion}
Our results shed light on the complex effects of The Great Ban, a paramount example of deplatforming that involved around 2,000 subreddits. Among our main findings is that 15.6\% of the affected users abandoned Reddit after the ban. Those who remained on the platform reduced their toxicity by 6.6\% on average. At the same time however, around 5\% of all users markedly increased their toxicity. The presence of such resentful users was widespread across the analyzed subreddits. These nuanced results cover new ground on the effects of The Great Ban and, more broadly, on adverse reactions to moderation interventions. Our results also lend themselves to multiple considerations about the design and deployment of effective moderation as well as about the challenges of regulating online platforms.

\textbf{Effectiveness of the moderation.} In literature, the effectiveness of content moderation actions has been primarily assessed in terms of changes that the moderation caused to the \textit{activity} and \textit{toxicity} of the affected users~\cite{chandrasekharan2017you,jhaver2021evaluating}. To this regard, our study revealed that a considerable share of toxic users abandoned the platform while the others exhibited a modest reduction in toxicity and a marked reduction in activity. At first glance, these results appear to be indicative of a successful moderation. However, it is necessary to deeply scrutinize these findings in light of potential unintended consequences. For example, the departure of some of the most toxic users from the platform suggests a form of \textit{displacement} rather than a resolution of the toxicity issue. As recent research pointed out, these users have likely migrated to other online spaces~\cite{horta2021platform}, raising concerns about the displacement of their toxic interactions instead of their mitigation. These worries are emphasized by the knowledge that users who migrate after facing restrictions on a platform, subsequently engage in even more toxic and aggressive behavior~\cite{horta2021platform}. In addition, user churn and diminished activity levels post-ban might pose challenges for Reddit, as online platforms thrive on user engagement and interactions for generating revenues~\cite{trujillo2022make}. Consequently, the apparent success of The Great Ban in mitigating toxicity must be interpreted with caution, considering the potential negative impact on the broader online ecosystem ---due to user migrations--- and on the platform's economic model ---due to user churn and reduced activity. Future endeavors should aim to strike a balance, devising strategies to curb toxicity without inducing abandonment or substantial decreases in user activity. The quest for effective moderation should align with the overarching goal of cultivating healthier online communities without compromising the safety of other platforms or economic viability, which could otherwise disincentivize platforms to carry out scrupulous moderation.

\textbf{Divergent reactions to moderation.} Our analysis also revealed that, in spite of a modest overall reduction in toxicity, a non-negligible minority of users exhibited large toxicity increases. This result has important implications for the assessment and development of moderation interventions. First, it sheds light on the complexity of user reactions to content moderation, which is a largely underexplored area that requires further investigation~\cite{jhaver2023personalizing}. In addition, it surfaces the need for a more nuanced and \textit{personalized} approach to content moderation. A generic intervention such as The Great Ban ---which involved thousands of subreddits and tens of thousands of users--- may not effectively address the diverse motivations and behaviors exhibited by the affected users. In our work, this was exemplified by the minority of resentful users who greatly increased their toxicity. Understanding the factors contributing to such divergent responses is paramount for developing effective moderation strategies. Future research and practical applications should delve into user profiling, considering individual characteristics, past behavior, and contextual factors to effectively tailor moderation interventions~\cite{cresci2022personalized}. Moreover, the migration of a subset of toxic users and the widespread presence of users experiencing heightened toxicity raises questions about the potential \textit{radicalization} effect of moderation and its unintended contribution to the amplification of echo chambers~\cite{horta2021platform}. It suggests that some users may react negatively to certain moderation actions, possibly leading to more extreme behaviors. These observations highlight the delicate balance required in content moderation, where the aim should not only be that of reducing toxicity locally, but also preventing inadvertent consequences that might exacerbate polarization or radicalization in some user groups.

\textbf{Limitations.} Our study is based on a large historical dataset of Reddit comments shared by 17K users during a long observation window of 14 months centered around The Great Ban. Due to the nature of our dataset, our findings may be specific to Reddit and the context of The Great Ban. Therefore, caution is needed in generalizing the results to other online platforms or different moderation interventions. Similarly, online platforms are dynamic environments that are subject to continuous changes in user behavior, community norms, and platform policies. Our study covers a specific snapshot in time, albeit relatively long, which nonetheless limits the possibility to carry over our results to different time periods and, partially, also to account for long-term or evolving trends.
Moreover, toxicity scores of comments were obtained via the machine learning model Detoxify, which in itself might have introduced a biased evaluation of toxicity.
Another limitation of our work lies in its observational nature, which hinders the possibility of accounting for external events or changes in the broader online ecosystem that may have influenced user behavior independently of The Great Ban. For this reason, care is needed in establishing causal relationships from the findings presented herein.
Future work could adopt more sophisticated causal inference techniques, for instance difference-in-differences or interrupted time series. However, finding suitable control subreddits and taking into account exogenous events remains a significant challenge~\cite{trujillo2022make}.
Finally, our dataset lacks comprehensive information about user demographics, motivations, or contextual factors. Accounting for these aspects could provide a more nuanced and actionable interpretation of our results. In this regard, future research could investigate the unexplored interplay between user characteristics and the outcomes of moderation, also as a preliminary step towards the development of targeted and personalized moderation interventions~\cite{cresci2022personalized}.

\textbf{Ethical considerations.} This research contributes to a deeper understanding of the impact of content moderation, shedding new light on the complexities of user reactions to moderation interventions. This knowledge can inform future developments of effective and nuanced moderation strategies aimed at curbing online toxicity while minimizing unintended consequences. 
To this end, our work draws attention on the trade off between common versus minority good. In our work this is exemplified by the ethical dilemma faced by moderators who must decide whether to enforce interventions that could possibly harm a minority of users by making them resentful, in order to provide a modest benefit to the broader community.

 \section{Conclusions}
\label{sec:conclusions}
The Great Ban was a massive deplatforming operation enforced to shut toxic communities on Reddit. To evaluate its effectiveness, we analyzed 16M comments shared over the course of 14 months by 17K users affected by the ban. Our results reveal that 15.6\% of the affected users abandoned Reddit after the ban and that those who remained reduced their toxicity by 6.6\% on average. Despite this modest toxicity reduction, 5\% of users increased their toxicity by more than 70\% of their pre-ban level. The presence of such resentful users was widespread across the analyzed subreddits rather than concentrated in a few ones.

Overall, our study provides new and nuanced insights into the effectiveness of The Great Ban, including its undesired consequences. As such, it can inform the development of future and more effective moderation interventions and the policing of online platforms. Specifically, future work could extend our present results by delving deeper into the relationship between user characteristics and the outcome of moderation interventions. In turn, this would pave the way to the development of targeted or personalized interventions that could mitigate the undesired effects of moderation actions, such as those discussed in this work. Other promising avenues of future research are the development of predictive models for the outcome of moderation interventions. These would allow to estimate the likely effects of an intervention in advance of its application, enhancing the possibility to plan the strategic enforcement of moderation actions.
 
\begin{acks}
This work is partially supported by the project \texttt{PIANO} (Personalized Interventions Against Online Toxicity) under CUP B53D23013290006 as part of the Italian PRIN 2022, and funded by th
e European Union's program \textit{Next Generation EU}; by the PNRR-M4C2 (PE00000013) "FAIR-Future Artificial Intelligence Research" - Spoke 1 "Human-centered AI", funded under \textit{Next Generation EU}; and by the Italian Ministry of Education and Research (MUR) in the framework of the \texttt{FoReLab} projects (Departments of Excellence).
\end{acks}

\balance
\bibliographystyle{ACM-Reference-Format}
\bibliography{references}

\end{document}